\documentclass[pra,amssymb,amsmath,twocolumn,floatfix]{revtex4}
\usepackage{graphicx}

\begin{document}

\title{Quantum Ratchets for Quantum Communication with Optical Superlattices}

\author{Oriol Romero-Isart}
\affiliation{Departament de F{\'\i}sica, Grup de F{\'\i}sica
Te\`orica, Universitat Aut\`onoma de Barcelona, E-08193
Bellaterra, Spain.}
\author{Juan Jos\'e Garc{\'\i}a-Ripoll}
\affiliation{Facultad de CC. F{\'\i}sicas, Universidad Complutense de Madrid,
  Ciudad Universitaria s/n, Madrid, E-28040, Spain.}

\begin{abstract}
  We propose to use a quantum ratchet to transport quantum information
  in a chain of atoms trapped in an optical superlattice.  The quantum
  ratchet is created by a continuous modulation of the optical
  superlattice which is periodic in time and in space. Though there is
  zero average force acting on the atoms, we show that indeed the
  ratchet effect permits atoms on even and odd sites to move along
  opposite directions.  By loading the optical lattice with two-level
  bosonic atoms, this scheme permits to perfectly transport a qubit or
  entangled state imprinted in one or more atoms to any desired
  position in the lattice. From the quantum computation point of view,
  the transport is achieved by a smooth concatenation of perfect swap
  gates. We analyze setups with noninteracting and interacting
  particles and in the latter case we use the tools of optimal control
  to design optimal modulations. We also discuss the feasibility of
  this method in current experiments.
\end{abstract}

\maketitle

\section{Introduction}

By a ratchet effect one usually refers to the existence of directed
transport in a system in which there is no net bias force. Ratchets
have been traditionally found in the study of dissipative systems
\cite{reimann97}, where external fluctuations causing Brownian motion
cooperate with a periodic force to bias transport, a mechanism that is
in the basis of some biological motors \cite{astumian97}. Ratchets can
also appear without dissipation. These Hamiltonian or conservative
ratchets are interesting as they can be extended to quantum mechanical
systems. In this context one finds studies that relate the existence
of transport to classical properties of the model, such as some
asymmetries of the external force \cite{flach00} or mixing of chaotic
and regular phase space regions \cite{schanz01}.

Ultracold atoms offer an ideal arena to test this phenomenology, as
can be seen from the experiments implementating both dissipative and
conservative ratchets
\cite{moore95,klappauf98,mennerat99,ringot00,jones04,behinaein06,gommers06,sanchez-palencia04,sjolund06,denisov07,poletti07}. These
experiments rely on the force imparted by near or far from resonance
laser beams which act on the atoms over short periods of time. These
flashing potentials implement variants of the $\delta$-kicked rotor
model and lead to phenomena such as dynamical localization.  On the
theoretical side we must remark the discovery of directed transport on
quantum mechanical systems whose classical counterpart is completely
chaotic \cite{monteiro02,jonckheere03,hutchings04,hur05, kenfack07}.

The present work aims at exploiting the fact that quantum ratchets can
be used to transport quantum states and thus distribute entanglement
between distant nodes, a basic task for quantum information and
computation purposes \cite{bennett00}.  Within this context, we find
two distinct research directions. On the one hand, entanglement
swapping and quantum teleportation \cite{bennett93} can be used to
build efficient quantum repeaters \cite{briegel98,duan01} for long
distance communication. On the other hand, on a smaller scale, a
relevant effort has been recently devoted to study quantum state
transfer using static local Hamiltonians which act on chains of spins,
harmonic oscillators, or bosons in optical lattices \cite{bose03,
  osborne04,amico04, christandl04,plenio04,wojcik05,haselgrove05,
  burgarth05, burgarth07,
  romero-isart07,eckert07,bayat07,kay07,damico07,romeroisart07b,bose07}.
Compared to these works, time-dependent (non-adiabatic) models, such
as the one presented here, provide more degrees of freedom to optimize
the efficiency, speed and robustness of the quantum transport.

\begin{figure}[t]
  \centering
  \includegraphics[width=\linewidth]{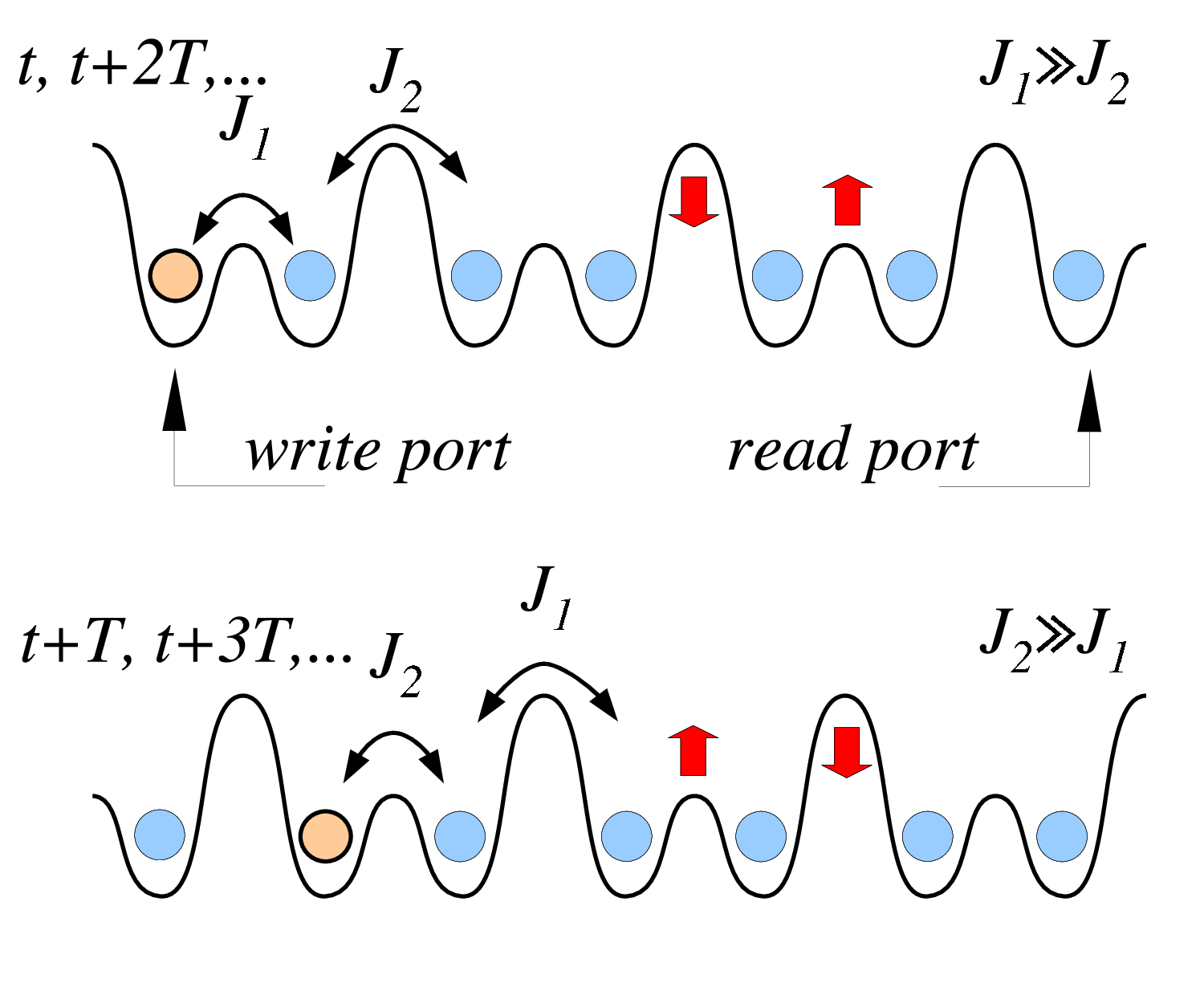}
  \caption{(Color online)An optical superlattice arises from a combination of two
    potentials with different periods. By modulating the depths and
    displacements of these potentials we can raise and lower the
    tunneling rates between odd and even pairs of sites (red
    arrows). This way, by means of perfect swaps, the state of a particle can be
    transported along the lattice.}
  \label{fig:superlattice}
\end{figure}

Moreover, the scheme we propose here for quantum state transport can
be seen as a quantum ratchet induced without breaking translational
and time reversal symmetries in the forces. The transport is just
triggered by the symmetry breaking of the initial state. We illustrate
this with a particular and simple implementation of quantum state
transport based on two-level bosonic ultracold atoms in optical
superlattices.  More specifically the transport is achieved as
follows. All the atoms are initially prepared in the, say, down
state. At a given site (write port) an arbitrary qubit state is
imprinted in the atom. The trapping potential is then modulated
smoothly and periodically in time and in space.  Depending on the
initial asymmetric position of the atom, being in an even or odd
lattice site, the qubit state is transported rightwards or leftwards
to any desired lattice site (read port) [Fig.~\ref{fig:superlattice}]
without precise individual addressing \footnote{During the completion
  of this manuscript, an article of C. E. Creffield
  {\cite{creffield07}} with related ideas has appeared. He shows how
  the quantum control of a periodic driving field can be employed to
  guide the motion of a boson in an otherwise empty optical lattice
  and to create entanglement by the interaction of two distinguishable
  bosons.}. We consider both free and interacting particles. To our
knowledge, this is the first time many-body interactions are taken
into account in Hamiltonian ratchets.

From the implementation point of view, this work is inspired by recent
advances in the control and manipulation of optical superlattices
\cite{foelling07,anderlini07,lee07,anderlini06,sebbystrabley06}.
Present experiments can create periodic potentials such as the ones
depicted in Fig.~\ref{fig:superlattice}, achieving a great control on
the time-dependence of the potential heights, and even being able to
measure the number of atoms on the even and odd sites
\cite{foelling07,anderlini07}. These tools suffice for the protocol
devised in this work.

The outline of the paper is as follows. In
Sect.~\ref{sec:superlattice} we introduce our system, made of a chain
of atoms in an optical superlattice. We state the goal of this work,
which is to transfer a quantum state between two arbitrary lattice
sites using a translationally invariant modulation of the
superlattice, and we present the Bose-Hubbard Hamiltonian that models
the dynamics of the atoms. In Sect.~\ref{sec:U=0} we study the case of
noninteracting particles and find that it is possible to induce
directed transport with arbitrary speed. In
Sect.~\ref{sec:interacting} we discuss a more realistic setup in which
atoms interact. We design a ratchet Hamiltonian with the help of
optimal quantum control, and find that we can still induce perfect
quantum transport with a speed limited by the interaction
strength. The details of the optimal quantum control technique are
left to Sect.~\ref{sec:control}. In Sect.~\ref{sec:experiments} we
study the optical superlattice modulation in more detail, discussing
the experimental challenges for implementing these quantum ratchets in
current experiments. Finally, in Sect.~\ref{sec:summary} we summarize
our results and comment on possible extensions.

\section{Optical superlattices}

\label{sec:superlattice}

\subsection{The model}

In this work we consider the setting of ultracold neutral atoms trapped in an optical
superlattice. An optical lattice is a standing wave of coherent
off-resonance light created by the interference of two or
more laser beams. As explained in Ref.~\cite{jaksch98}, such standing
wave behaves as a periodic potential that confines the atom in the
minima or maxima of intensity. The optical superlattice arises from a combination of two potentials with different periods. In the simplest case the superlattice
potential can be written as follows
\begin{eqnarray}
  \label{superlattice}
  V(x,y,z,t) &=& V_{x}(t)\cos^2(kx) + V_2(t) \cos^2 (2kx+\phi) +\nonumber\\
  &+& V_{\perp}[\cos^2(ky)+\cos^2(kz)].
\end{eqnarray}
Here $V_x(t)$ and $V_\perp$ are the
strengths of the lattice along the main axis and transversely to it, and
$k=2\pi/\lambda$ is the momentum of the photons. In order to have a configuration of decoupled 1D lattices  \cite{greiner02}, the transverse potential  $V_\perp$ has to be much
larger than the recoil energy $E_r=\hbar^2k^2/2m$ to prevent tunneling
between 1D lattices. On top of this potential and along the main axis we find another lattice
of strength $V_2(t)$, with half the period of the original one and a
possible dephasing $\phi\in \mathbb{Z} \times \pi/2$. This yields to the optical superlattice that creates the double-well structure shown in
Fig.~\ref{fig:superlattice}.

We are interested in the regime of deep lattices in which atoms are
confined to the lowest Bloch band of the periodic potential. We assume
the atoms to be bosons with two internal states. Under those
circumstances the dynamics of the atoms can be modeled using a
Bose-Hubbard Hamiltonian \cite{jaksch98} which, for a single 1D tube
of lattices, has the form
\begin{equation}
  \label{model}
  H = \sum_{\sigma=\uparrow,\downarrow}\sum_{i=1}^L \left[-
    J_i (a_{\sigma i}^\dagger a_{\sigma i+1} + \mathrm{H.c.})
  + \frac{U}{2}a_{\sigma i}^{\dagger 2} a_{\sigma i}^2\right].
\end{equation}
Here, $a_{\sigma i}$ and $a_{\sigma
  i}^\dagger$ are Fock operators that annihilate and create particles on
the $i$-th site of the lattice and in one of two internal states,
$\sigma \in \{\uparrow,\downarrow\}$. The parameter $U$ is the on-site
interaction between atoms, which we assume to be spin independent, and
$J_i$ is the tunneling amplitude, which may vary from site to site.

In order to describe the dynamics of the atoms, Eq.~(\ref{model}) has
to satisfy two conditions. First of all, the tunneling $J_i$ between
different wells has to be small, so that there effectively exists a
low energy band formed by the localized states on each site. This
implies that the superlattice cannot drop the barrier between pairs of
sites too much [Fig.~\ref{fig:superlattice}], for otherwise we would
have to use a multiband model. The second condition is that the
on-site interaction $U$ must be small compared to the energy gap
between Bloch bands. Both condition are easily met in current
experiments with optical superlattices
\cite{lee07,anderlini06,sebbystrabley06,foelling07}, and have been
taken into account throughout this work [See
Sect.~\ref{sec:experiments}].

\subsection{The goal: quantum communication}
\label{sec:goals}

Our objective is as follows. We initially prepare all the two-level atoms in the same state, say the down state $\left|\downarrow\right\rangle$. We imprint an arbitrary quantum state on a particular site of the lattice, the \textit{write port}, in such a way
that the atom ends up in a superposition of both spin states,
$\alpha \left|\uparrow\right\rangle + \beta
\left|\downarrow\right\rangle$. The goal is  to create, by modulating the energy barriers of
the optical superlattice,  a ratchet effect that will
transport this state a predetermined distance, to another site, the
\textit{read port} in Fig.~\ref{fig:superlattice}. The modulation of
the superlattice will not break the translational invariance and it
will also be periodic in time (with a period equal to $2T$),
alternating the roles of the hopping between even ($J_{2n}=J_2$) and odd ($J_{2n+1}=J_1$) sites
\begin{equation}
  \label{periodicity}
  J_{1,2}(t+T) = J_{2,1}(t).
\end{equation}

The Hamiltonian in Eq.~(\ref{model}), being translationaly  invariant
in time  and in space, does not exert any force on the atoms, as
expected from a quantum ratchet. This can be verified by computing the
integral of the forces acting on an atom in a well of the lattice. In
particular, the following average is zero
\begin{equation}
  \label{eq:1}
  \int_0^{2T}dt \int_{0}^{\frac{\pi}{2k}}dx \frac{d}{dx}V(x,y,z,t) = 0.
\end{equation}
Note however that our superlattice does not have a permanent asymmetry
on the unit cell $[0,\pi/2k)$, unlike other Hamiltonian flashed
ratchets with saw-tooth profiles \cite{mennerat99}.

While the average force acting on the atoms is zero, we will show that
there is still transport.  The reason is the \textit{asymmetry of the
  initial state}
\begin{equation}
  \label{state}
  |\psi(0)\rangle =
  (\alpha a_{\uparrow \text{in}}^\dagger + \beta a_{\downarrow \text{in}}^\dagger)
  \prod_{j \neq \text{in}} a_{\downarrow j}^\dagger |\mathrm{vac}\rangle,
\end{equation}
and the fact that our time-dependent Hamiltonian will perform perfect
swaps between neighboring lattice sites. All of this leads, after $n$ swaps of duration $T$, to a
propagation of the imprinted qubit rightwards or leftwards
\begin{equation}
  |\psi(n T)\rangle =
  (\alpha a_{\uparrow \text{out}}^\dagger + \beta a_{\downarrow \text{out}}^\dagger)
  \prod_{j \neq \text{out}} a_{\downarrow j}^\dagger |\mathrm{vac}\rangle,
\end{equation}
depending on the starting site. Note also that by design, we will
achieve the transport of the quantum state without actually moving any
atom, thus leaving the number of atoms per site invariant.

Finally, let us mention that this scheme can also be used to
distribute entanglement in the same fashion as in \cite{amico04}.
Namely by imprinting a maximally entangled state in two
neighboring lattice sites. Since atoms in odd and even sites move on
opposite directions, the atoms will depart from each other and
entanglement will get distributed between arbitrarily distant lattice
sites.

\section{Noninteracting case}

\label{sec:U=0}

We start with the case of noninteracting particles $U=0$.
This case is relatively easy to study
as all particles can be treated independently and therefore we can focus on the dynamics of a single particle that starts
from different sites, a problem which is integrable.  The basis of
states is now denoted by $|j\rangle := a_j^\dagger
|\mathrm{vac}\rangle$, where we no longer care about the internal
state of the particle. Our only concern is now to ensure that the
state with a particle on the write port $|j=\mathrm{in}\rangle$ is mapped after
a given time, $t_r$, to a state with that particle on site
$|j=\mathrm{out}\rangle$, the reading port, in a deterministic fashion.  Note
also that since particles do not interact we do not have to consider the
dynamics of particles on other lattice sites.  
Nevertheless, as we
show below, it is possible to look for solutions that do not give
energy to these other particles and leave the density profile
invariant. If the number of atoms per lattice site is constant, we can
then talk about transport of the quantum state and not of particles
themselves [Fig.~\ref{fig:superlattice}].

\begin{figure}[h!]
  \centering
  \includegraphics[width=0.9\linewidth]{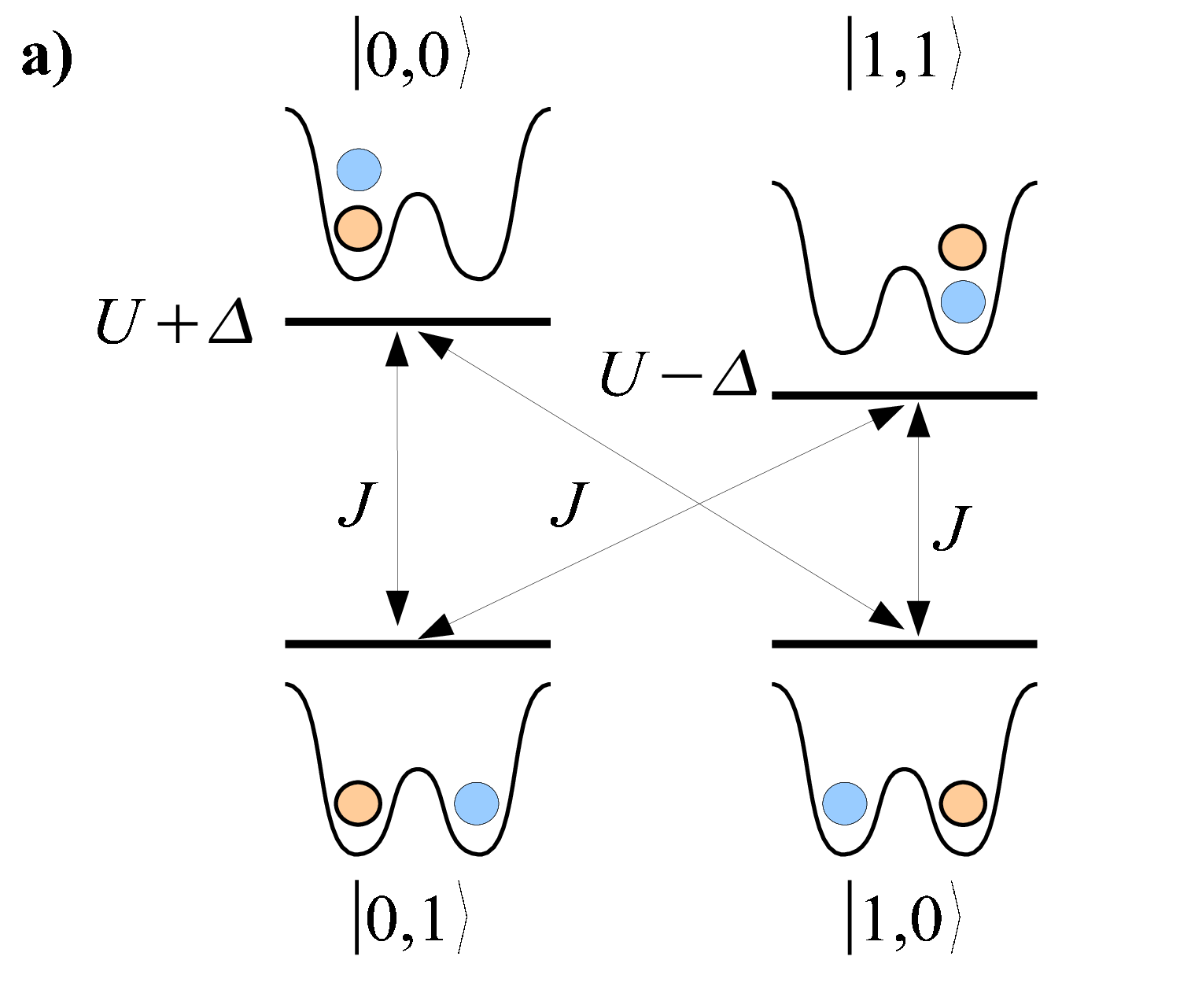}
  \includegraphics[width=0.9\linewidth]{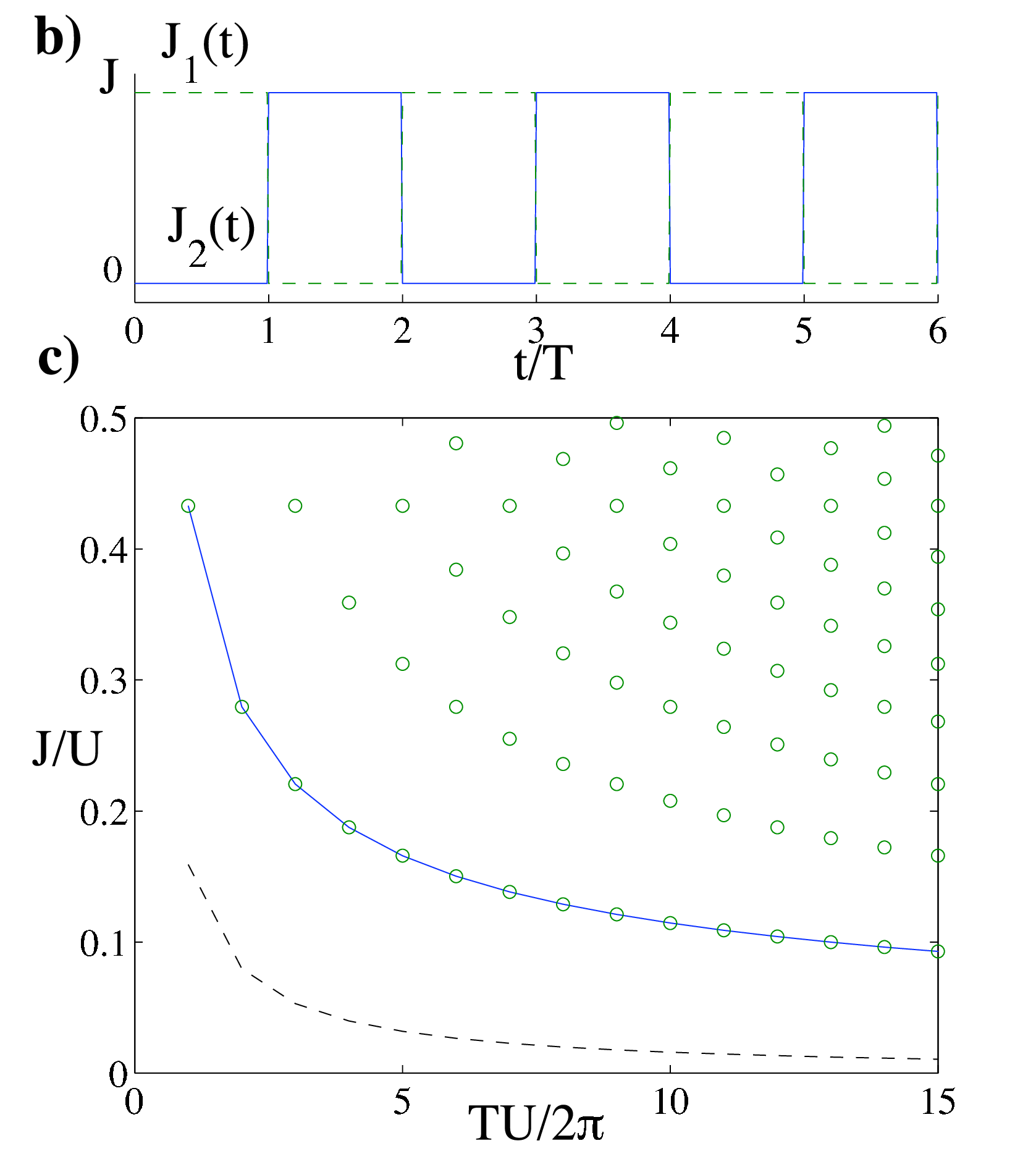}
  \caption{(Color online) (a) Model case of two lattice sites disconnected from the
    rest because $J_2=0$. By fixing the hopping between wells to a
    precise value $J$ for a time $T$, it is possible to swap the
    atoms. Note that the doubly occupied sites have more energy due to
    the interaction, $U$, and a possible inhomogeneity of the lattice
    potential, $\Delta$.  (b) Combining the solution for a pair of
    sites we obtain one possible modulation of the hoppings $J_1$
    (solid) and $J_2$ (dashed), namely $J_1$ and $J_2$ are square waves in antiphase. This produces     the quantum transport. (c) Values of $J$ and $T$ for which perfect transport
    is achieved. Each circle is a solution; the solid line joins the
    solutions with smallest hopping and the dashed line is the
    solution for noninteracting particles.}
  \label{fig:3level}
\end{figure}

As described before, we will adopt two restrictions. The first one is
that the couplings do not break the translational invariance but are
modulated in time, with the roles of $J_2$ and $J_1$ being exchanged
after a time $T$, as from Eq. (\ref{periodicity}). The second one is to assume that
the tunneling can be brought to zero, so that
\begin{eqnarray}
  J_1(t) \geq 0,\; J_2(t) &=& 0,\quad t\in [0,T),\nonumber\\
  J_{1,2}(t+T) &:=& J_{2,1}(t).
\end{eqnarray}
During the time in which $J_2=0$ we have ($\hbar=1$)
\begin{eqnarray}
  \left(\begin{array}{c}a_{2n+1}(t) \\ a_{2n+2}(t)\end{array}\right) =
  \{\cos[\theta(t)] \mathbb{I}+ i \sin[\theta(t)] \sigma^x\}
  \left(\begin{array}{c}a_{2n+1}(0) \\ a_{2n+2}(0)\end{array}\right),  
\end{eqnarray}
where $\sigma^x$ is a Pauli matrix and $\theta(t)=\int_0^t
J_1(\tau)d\tau$. The simplest solution of this kind is a square
signal, $J_1(t)=J$, as shown in Fig.~\ref{fig:3level}b.  This means
that for $T = \pi/(2J)$ we achieve a perfect swap between even and odd
sites up to an irrelevant global phase. In practice the tunnelings $J_1$ and $J_2$ will evolve smoothly
and require some time to reach a nonzero value. In that case the time
for perfect switch will be given by $\theta(T) = \pi/2$.

It is possible to construct arbitrarily fast solutions of also
arbitrary smoothness. Note, however, that the value of the average
hopping is inversely proportional to the modulation period, $\bar J
\propto 1/T$, so that a fast solution may require a lattice with an
unrealistically large hopping.


\section{Interacting case}

\label{sec:interacting}

The case of interacting particles is more realistic but also more
difficult. We can no longer regard the particles as independent and
collisions can affect the phase of the transported state, as defined by Eq. 
(\ref{state}). We will still look for simple solutions that
concatenate a swap dynamics on pairs of lattice sites, with the
restrictions introduced in Sec.~\ref{sec:goals}.

\subsection{A two-level problem}

Let us repeat the calculation of Sect.~\ref{sec:U=0} but now taking
into account the interaction between particles. Since $J_2=0$ we focus
on the double-well problem with only two distinguishable
particles, {\em i.e.} in two different internal states. Instead of using second quantization we denote their state
as $|i\rangle|j\rangle$, where $i,j\in\{0,1\}$ is the well in which
the particle resides. The connections between states are depicted in
Fig.~\ref{fig:3level}a. Using Pauli operators the effective
Hamiltonian becomes \footnote{We will assume that the inhomogenity
  $\Delta$ in Fig.~\ref{fig:3level} is zero, because otherwise the
  lattice has a net bias force and we cannot talk about a
  ``ratchet''.}
\begin{equation}
  H = -J (\sigma^x\otimes\mathbb{I}+\mathbb{I}\otimes\sigma^x)
  + \frac{U}{2}(\sigma^z\otimes\sigma^z+1)
\end{equation}
We can further restrict our problem to the only states that
participate on the dynamics
\begin{eqnarray}
  |\psi^- \rangle &:=& \frac{1}{\sqrt{2}}(|01\rangle - |10\rangle),\\
  |\psi^+ \rangle &:=& \frac{1}{\sqrt{2}}(|01\rangle + |10\rangle),\nonumber\\
  |\phi^+ \rangle &:=& \frac{1}{\sqrt{2}}(|00\rangle + |11\rangle).\nonumber
\end{eqnarray}
Using this basis we obtain the effective Hamiltonian
\begin{equation}
  \label{eq:2}
  H(t) = \left(
    \begin{array}{ccc}
      0 & 0 & 0 \\
      0 & 0 & -2J(t) \\
      0 & -2J(t) & U
    \end{array}\right),
\end{equation}
where $|\psi^-\rangle$ is shown to be a dark state and the other two
are coupled by a two-level Hamiltonian $2J(t)\sigma^x+U\sigma^z$. Our
problem is thus to find a hopping $J(t)$ such that after a time $T$
the states above have experienced the following transformation
\begin{eqnarray}
  |\psi^-\rangle \to |\psi^-\rangle,\; |\psi^+\rangle \to -|\psi^+\rangle,\;
  |\phi^+\rangle \to e^{i\nu}|\phi^+\rangle,
\end{eqnarray}
where the phase $\nu$ is unimportant for our purposes.

\subsection{The square signal revisited}

We will first investigate solutions which are piecewise constant, with
$J_1(t) = J,$ for $t\in[0,T)$. For a fixed hopping our Hamiltonian has
two nonzero eigenvalues
\begin{equation}
  E_{\pm} = \frac{U \pm \sqrt{16J^2+U^2}}{2}.
\end{equation}
that contribute to the evolution of the symmetric states. The perfect
swap between the atoms takes place at a time $T$ such that the
symmetric state $|\psi^+ \rangle $ changes sign. As $|\psi^+ \rangle $ can be written as a superposition of the two symmetric eigenstates with energies $E_\pm$, this condition is satisfied when $E_{\pm} T = \pm
(2n_{\pm}+1)\pi$, where $n_{+}$ and $n_{-}$ are arbitrary
integers. Defining $x := (2n_++1)/(2n_-+1)$ we obtain
\begin{equation}
  \frac{U}{J} = 2\frac{x-1}{\sqrt{x}}.
\end{equation}
From this value we can compute the energies $E_{\pm}$ and the time
$T$.  As shown in Fig.~\ref{fig:3level}c the minimal value of $J$ is
reached for either $n_+$ or $n_-$ equal to 0 and, unlike in the
noninteracting case, there exists a minimal swap time given by
$T=2\pi/U$. Indeed, using the tools in \cite{khaneja01} one can prove
that the constant hopping is the fastest solution and that the gate
cannot be performed faster than this time.

\subsection{Smooth solutions: optimal quantum control}

\begin{figure}
  \centering
  \includegraphics[width=\linewidth]{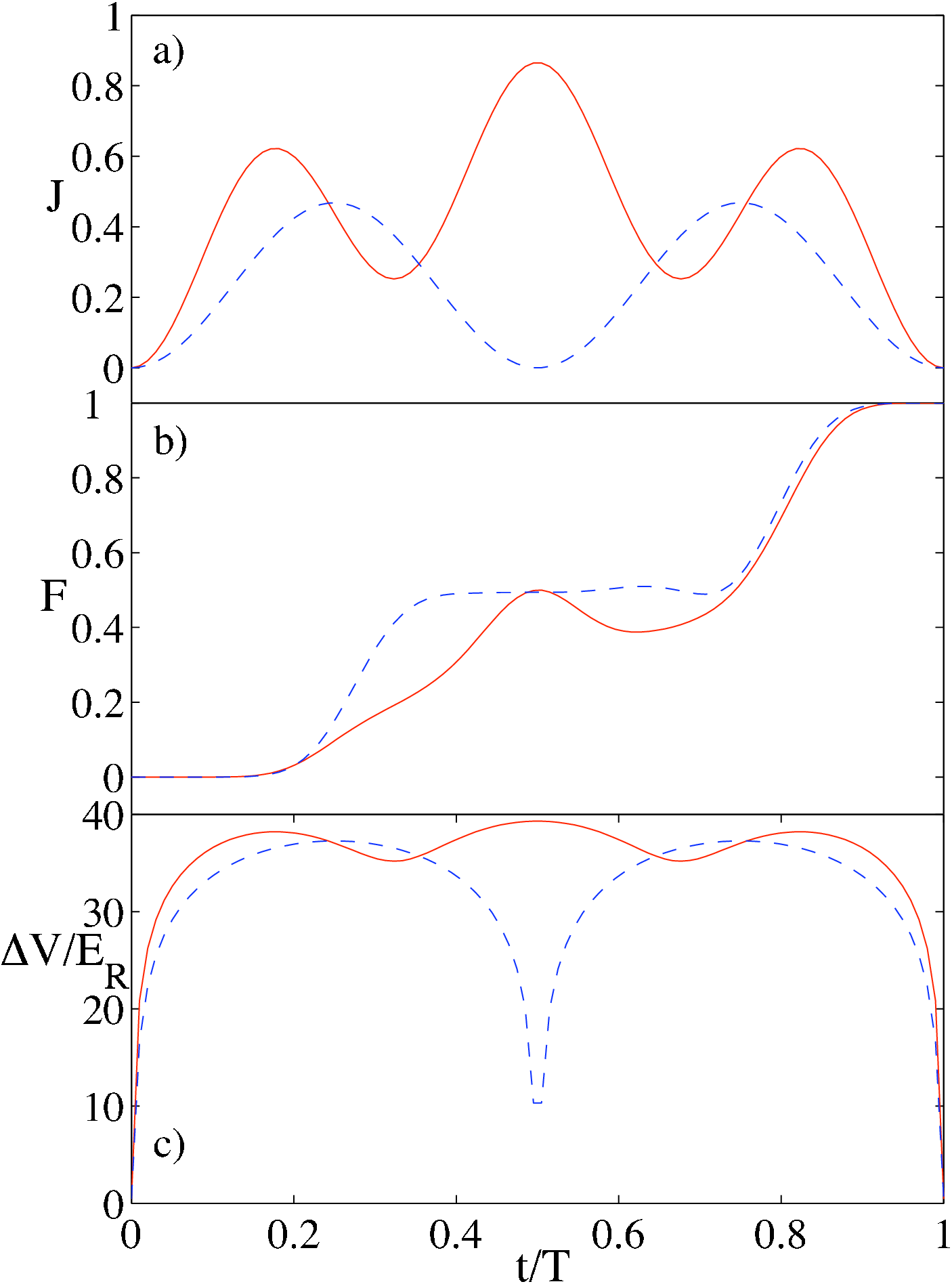}
  \caption{(Color online)(a) Hopping and (b) fidelity of the gate ($F=|\langle 01 |U(t)|10\rangle|^2$) with the perfect
    transport for half a period $[0,T)$ during which $J_1(t)=J(t)$ and
    $J_2=0$. We find two solutions, one for $T=2\pi/U$ (solid) and
    another one for $T=4\pi/U$ (dashed), with three and two modes,
    respectively. In (c) we plot the corresponding modulations of the
    superlattice.}
  \label{fig:fidelity}
\end{figure}

While optimal, the square signal that we have considered before is
probably unrealistic, as in experiments the hopping will smoothly
increase and decrease as the tunneling barriers are changed. For that
reason we have investigated other solutions with continuous
derivatives using the tools of optimal quantum control.

The details of the method are left for Sect.~\ref{sec:control}, but
let us sketch the procedure. The first step is to parametrize the
hopping as a linear combination of some functions
\begin{equation}
  J(t) = \sum_n c_n f_n(t) \geq 0
\end{equation}
where for simplicity we use trigonometric functions $f_n(t) = \sin^2(\pi
n t / T)$ that increase and decrease smoothly to zero. Since the
hopping is positive we have a first restriction, $c_n \geq 0$. The
second restriction is that the fidelity of the swap procedure has to be
one. Both constraints are imposed to the problem of minimizing the
average strength of the hopping, given by $E=\sum_n |c_n|^2$. This
problem is solved numerically with MATLAB's optimization
toolbox \cite{branch99} using as aid the derivatives computed by means
of perturbation theory [Sect.~\ref{sec:control}].

In Fig.~\ref{fig:fidelity} we show two instances of the problem, one
optimized for three modes ($c_n=0$ for $n>3$) and a duration of $T=2\pi/U$, the other one
for twice the time and two modes. As it can be appreciated in the picture, we reach
perfect fidelity in a rather smooth and robust manner, so that errors
in the timing of the gate will not affect the process significantly.

\begin{figure}
  \centering
  \includegraphics[width=0.9\linewidth]{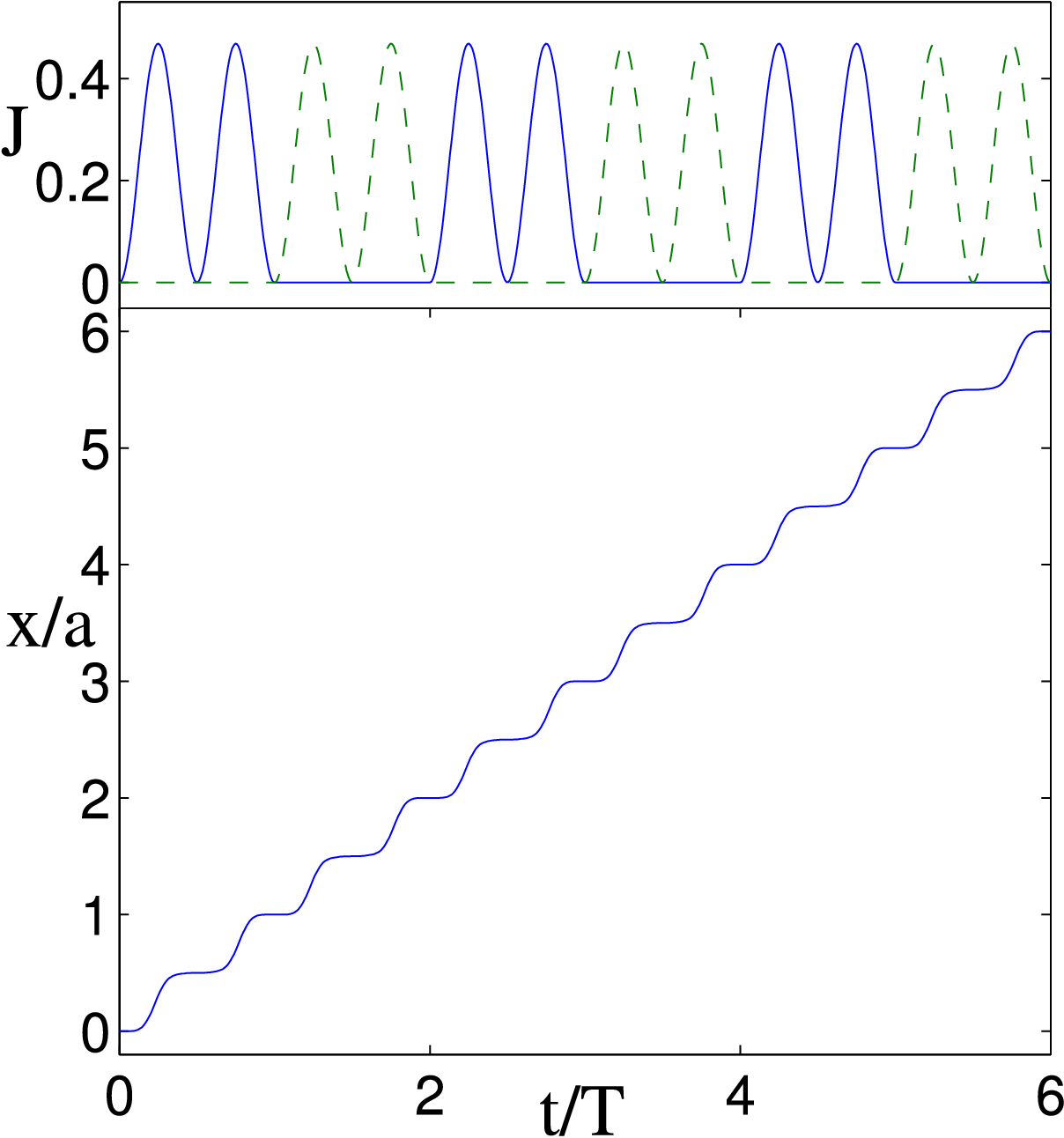}
  \caption{(Color online)On the upper figure we plot a solution of $J_1(t)$ (solid)
    and $J_2(t)$ (dashed) for a perfect transport of the qubit. In the
    lower figure we plot the average position of the qubit on the
    lattice as transported by these modulations. These plots have been
    computed using $T=4\pi/U$.}
  \label{fig:transport}
\end{figure}

To further relate these solutions to the notion of a ratchet let us
look at Fig.~\ref{fig:transport}. There we plot the full time
evolution of the imprinted qubit state for 6 periods of an optimal
modulation. The position may oscillate between pairs
wells, but there is always a net average transport.

\subsection{Dealing with holes}

The two solutions derived above, that is the piecewise constant and
the smooth ones, are designed to induce transport on a chain of
particles. However, in practice such a chain will have some endpoints
or particles that stand near an empty site. We can then have three
scenarios: (i) that a particle standing near a hole ends up in the
original site at time $t=T$; (ii) that the particle and the hole are
swapped and, more generally, (iii) that the particle and the hole end
up in some coherent superposition of being on each site. Out of these
processes, only the latter will affect the evolution of the state we
want to transport, since there is a small probability that it gets
reflected.

Holes only have a disturbing effect on the transport if $U\neq 0$,
since in the noninteracting case surrounding particles are equivalent
to holes. Note however that we can effectively eliminate the scenario
(iii) if we impose that the hopping on half a period leaves the
particle invariant
\begin{equation}
  \int_0^T J(\tau) d\tau = 2\pi \times \mathbb{Z},
\end{equation}
While this restriction cannot always be achieved for the piecewise
constant profile, it can be easily incorporated to our optimal control
toolbox.

\section{Experimental implementation}

\label{sec:experiments}

\begin{figure}
  \centering
  \includegraphics[width=0.9\linewidth]{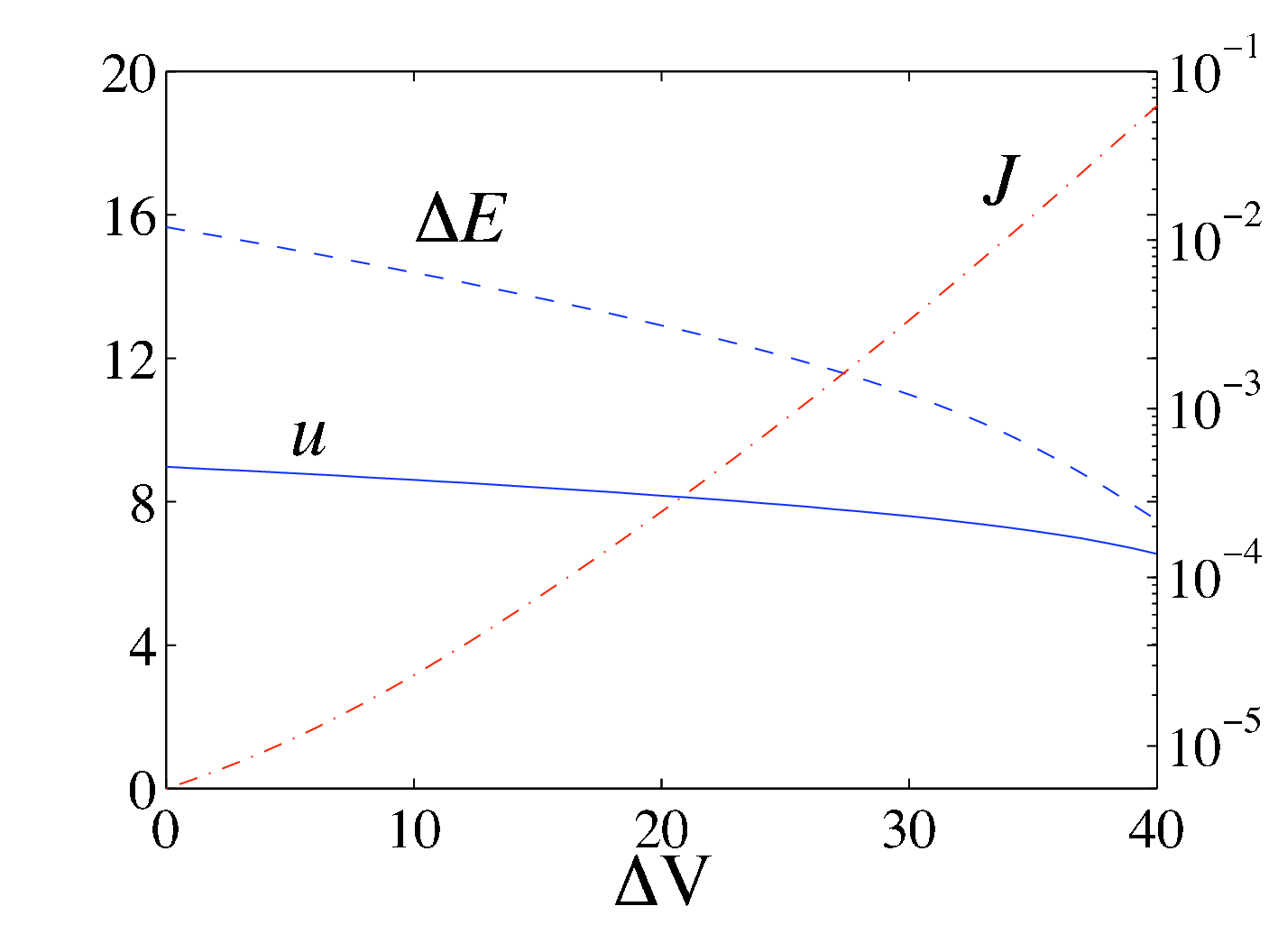}
  \caption{(Color online)Parameters of the double well potential as a function of
    the superlattice modulation, $\Delta V = (V_x + V_2)/2$. We plot
    the effective hopping between wells, $J$ (dash-dot, right axis),
    the energy gap to higher bands, $\Delta E$ (dashed), and the
    on-site interaction, $u$ (solid). Everything is expressed in units
    of the recoil energy, $E_r$.}
  \label{fig:wannier}
\end{figure}

All the protocols that we have designed can be implemented in current
experiments with optical superlattices
\cite{anderlini07,foelling07,lee07,anderlini06,sebbystrabley06}. The
implementation should begin by loading the superlattice with
approximately one atom per site, all of them in the same internal
state. The next step is to rotate the state of some atoms using either
magnetic field gradients, coherent light or a clever combination of
both \cite{foelling06}, in order to imprint the quantum state we wish
to transport. By modulating in time the intensity of the laser beams
that participate in the optical lattice one can then achieve a
modulation of $J_1(t)$ and $J_2(t)$ that corresponds to the solutions
studied above. After an appropiate time one may retrieve the qubit
state by measuring the lattice sites on which it is expected to
arrive.

In an experiment one does not directly control the hoppings $J_1(t)$
and $J_2(t)$, but rather the lattice strengths $V_x$ and $V_2$ [See
Eq.~(\ref{superlattice})]. In order to relate these two quantities we
have have performed a band structure calculation. We have focused on
the case $J_2(t)\simeq 0$, which corresponds to the first half-period,
$t\in[0,T)$. This hopping can be suppressed by making $\phi=0$ and
$V_x+ V_2 = 70 E_r$, which is large enough to effectively suppress
hopping every second site. Given this constraint, any modulation of
the lattices depends on a single quantitiy, $\Delta V(t) \geq 0$, such
that $V_x = \Delta V(t)$ and $V_2 = 70E_r - \Delta V(t)$. It thus
remains to relate $\Delta V(t)$ and $J_1(t)$. This is done for each
possible value of $\Delta V$, by computing numerically the ground
state wavefunction of a particle in a double-well, $w(x)$, and the
first excited state. The energy difference between these states is
proportional to the effective tunneling $J_1(t)$, while the on-site
interaction energy becomes
\begin{equation}
  \label{eq:3}
  \frac{U}{E_r} = \frac{a_s}{a} u,
\end{equation}
where $a_s$ is the scattering length between atoms, $a$ the period of
the superlattice and $u = \int |w(x)|^4$.

The results are shown in Fig.~\ref{fig:wannier}. Since $a_s/a$ is very
small, the on-site interaction energy is smaller than the energy gap
between Bloch bands and the Bose-Hubbard model is therefore valid
throughout the evolution (\ref{model}). Furthermore, $u$ does not
change very much as a function of the modulation $\Delta V$, while $J$
decays exponentially fast with $\Delta V$. After an appropiate fit of
these quantities, one may convert the time dependence of $J_1(t)$ from
Fig.~\ref{fig:fidelity}a into the associated modulations of the
superlattice $\Delta V(t)$, which are shown in
Fig.~\ref{fig:fidelity}c.

\section{Coherent control}

\label{sec:control}

In this section we present the tools that we have used to
optimize the modulation of the hopping, as they differ significantly
from what is the standard approach in optimal quantum control based on a Lagrangian formulation \cite{werschnik07}.

\subsection{Objective function}

Let us formulate our problem: we have a Hamiltonian, $H(t;x)$
that depends both on time and on some additional parameters,
$x_1\ldots x_M$.  Our goal is to find an optimal set of parameters of the Hamiltonian such that the evolution of a number of states is as equivalent as possible to a transformation given by a specific unitary $U_g$. 

Stated in a more concrete way, the evolution of an arbitrary state
$|\psi(0)\rangle$ is given by $|\psi(t)\rangle = U(t;x)
|\psi(0)\rangle$, where the unitary is a solution of the Schr\"odinger
equation
\begin{equation}
  i \frac{d}{dt} U(t;x) = H(t;x) U(t;x) \label{schrod}
\end{equation}
with initial condition $U(0;x) = \mathbb{I}$ \footnote{Note that we
  only specify the final time, but assume that the initial time is set
  at $t=0$.}. Our goal is to maximize the fidelity (defined below) of any evolved
state $U(t;x)|\psi_n\rangle$ with the desired transformed state
$U_g|\psi_n\rangle$. There are many ways to measure the accuracy of
the transformation. A strict and simple objective function is the fidelity
\begin{equation} \label{eq:fid}
   F = \frac{1}{d} \mathrm{Re}\{\mathrm{tr}[U_g^\dagger U(T;x)]\}
  = \frac{1}{d} \mathrm{Re} \sum_{n=1}^d \langle \psi_n | U_{g}^\dagger
  U(T;x) | \psi_n\rangle,
\end{equation}
where the $\{ | \psi_n \rangle \}$ form an orthonormal basis of the
$d$-dimensional Hilbert space ${\cal H}$ where we want to control the
evolution \footnote{It is also possible to perform the optimization
  over a smaller subspace, using fewer vectors. This is
  particularly useful when one wants, for instance, to control the
  evolution of a single state.}.  This function is bounded by 
$F \in [-1,1]$ and achieves the maximum value for the perfect transformation,
\begin{equation}
  F = 1 \,\,\Leftrightarrow\,\, U(T;x) = U_g.
\end{equation}
The question is thus, how do we maximize $F$? A natural way is to compute (when possible) the derivative of $F$ with respect to the parameters
$x$, i.e.  $\partial F / \partial x_i$, since the gradient itself
provides a direction along which the fidelity is increased. Indeed,
given this derivative there are multiple optimization algorithms that
would allow us to compute the optimal control.

\subsection{Formal gradient}

In order to obtain the gradient of $F$ with respect to the parameters $x$, we straightforwardly  obtain from \eqref{eq:fid}
\begin{eqnarray}
  \frac{\partial F}{\partial x_i} =
  \frac{1}{d} \mathrm{Re} \sum_n
  \left \langle \psi_n \left |U^\dagger_g \frac{\partial}{\partial x_i}
      U(T;x) \right |\psi_n \right\rangle,
  \label{fidelity}
\end{eqnarray}
which relates the gradient of $F$ to a derivative of the unitary
operator.  What follows is a simple way to compute $\partial
U/\partial x_i$ which is based on performing a Taylor expansion of the
operator $U(t;x)$ with respect to the parameters $x$
\begin{equation}
  \label{taylor}
  U(t;x+\epsilon) =
  U(t;x) + \epsilon \frac{\partial U}{\partial x}(t;x) 
  + {\cal O}(\epsilon^2).
\end{equation}
We will obtain this series using time-dependent perturbation theory on
the Schr\"odinger equation (\ref{schrod}), which will enable us to identify the derivative of the unitary operator.

Let us assume that by changing $x \to x+\epsilon$ our Hamiltonian
decomposes into an unperturbed part, $H_0=H(t;x)$ and a perturbation $H_1$
\begin{equation}
 \epsilon H_1 \equiv H(t;x+\epsilon) - H_0 =
 \epsilon \frac{\partial H}{\partial x}(t;x) + {\cal O}(\epsilon^2).
\end{equation}
The new unitary operator will satisfy a Schro\"odinger equation with a
modified Hamiltonian
\begin{equation}
  i \frac{d}{dt} U(t;x+\epsilon) = (H_0+\epsilon H_1)
  U(t;x+\epsilon),
\end{equation}
and same initial condition $U(0;x+\epsilon)=\mathbb{I}.$ It is
now convenient to move to the interaction picture
\begin{equation}
  U(t;x+\epsilon) \equiv U(t;x) W(t;x),\label{intpict}
\end{equation}
which leads to a simpler equation
\begin{equation}
  i \frac{d}{dt} W(t;x) =\epsilon
  U(t;x)^\dagger H_1(t;x) U(t;x) W(t;x).
\end{equation}
Integrating formally this differential equation and iterating the
resulting formula, one obtains the usual Dyson series. We only need
this series up to first order
\begin{equation}\label{eq:dyson}
  W(t;x) = \mathbb{I} - i \epsilon \int_0^t d\tau
  U(\tau;x)^\dagger H_1(t;x)U(\tau;x) + {\cal O}(\epsilon^2).
\end{equation}
From (\ref{intpict})
combined with \eqref{eq:dyson} we have
\begin{eqnarray}
  \label{perturbation}
  U(t;x+\epsilon) &=& U(t;x) - \\ &-&  i \epsilon U(t) \int_0^t d\tau
  U(\tau)^\dagger H_1(\tau) U(\tau) + {\cal O}(\epsilon^2)\nonumber
\end{eqnarray}
(In the second term and hereafter we omit the $x$ dependence to ease the notation).
We can now compare this expression with \eqref{taylor} in order to identify the derivative of the unitary, that is
\begin{equation}
  \label{derivative}
  \frac{\partial}{\partial x}U(t) = - i \epsilon  U(t)\int_0^t d\tau
  U(\tau)^\dagger \frac{\partial H}{\partial x}(\tau)
  U(\tau).
\end{equation}

Therefore we obtain the formula for the gradient of the fidelity
\begin{equation}
  \frac{\partial F}{\partial x_i} =\frac{1}{d}
  \mathrm{Re} 
  \sum_n
  \int_0^T d\tau
  \left \langle U_{g}^\dagger U(T)
  U(\tau)^\dagger \frac{\partial H}{\partial x_i}(\tau) U(\tau)
  \right\rangle_{\psi_n}.
\label{final-derivative}
\end{equation}

\subsection{Development of the algorithm}

Even though we have a closed expression for the derivative
of the fidelity, we still need to compute the
integral which appears in Eq.~(\ref{final-derivative}). We have
devised a simple, accurate and efficient procedure which is based on
solving three sets of ordinary differential equations.  The first one
is obtained by transforming the integral in
Eq.~(\ref{final-derivative}) into $d\times M$ ordinary differential
equation
\begin{equation}
  \frac{d}{dt}f_{n,i}(t)  = 
  \frac{1}{d}\mathrm{Re} 
  \left \langle \psi_n \left| U_{g}^\dagger
  U(T)U(t)^\dagger \frac{\partial H}{\partial x_i}(t) U(t)
  \right|\psi_n \right\rangle
\label{aux1}
\end{equation}
with initial condition $f_{n,i}(0) =0$ and final value
\begin{equation}
  \label{deriv-c}
  \frac{\partial F}{\partial x_i} = \sum_n f_{n,i}(T).
\end{equation}
We now notice that left side of the scalar product in Eq.~(\ref{aux1})
is the state $U(t) U(T)^\dagger U_g |\psi_n\rangle$ and can be
computed by solving an ordinary differential equation
\begin{equation}
  i \frac{d}{dt} |\xi_n(t)\rangle = H(t) |\xi_n(t)\rangle,
  \label{reverse}
\end{equation}
with initial condition $|\xi_n(T)\rangle = U_g|\psi_n\rangle$ and
moving backwards in time from $T$ to $t$. 

We now have all the ingredients to design the protocol that computes
our derivative (\ref{final-derivative}). We summarize it as
follows. First, solve Eq.~(\ref{reverse}) backwards in time up to
$t=0$, finding
\begin{equation}
  |\xi_n(0)\rangle := U(T)^\dagger U_{g} |\psi_n\rangle.
\end{equation}
We then solve the system of ordinary differential equations
\begin{subequations}
\begin{eqnarray}
  i\frac{d}{dt}|\psi_n(t)\rangle &:=& H(t) |\psi_n(t)\rangle\\
  i\frac{d}{dt}|\xi_n(t)\rangle &:=& H(t) |\xi_n(t)\rangle\\
  \frac{d}{dt}f_{n,i} &:=& \frac{1}{d}\mathrm{Im}
  \langle \xi_n(t) | \frac{\partial H}{\partial x_i}(t) |\psi_n(t)\rangle.
\end{eqnarray}
\end{subequations}
With the initial conditions $|\psi_n(0)\rangle$ being some orthonormal
basis of our Hilbert space, $|\xi_n(0)\rangle$ computed before and
$f_{n,i}(0)=0$. The value of the derivative is then computed using
Eq.~(\ref{deriv-c}).

This derivative and the formulas given above can be fed to any
optimization package, such as a simple line search algorithm or a
nonlinear conjugate gradient method. In particular we have used
Matlab's nonlinear optimization toolbox \cite{branch99}.

\section{Conclusions}

\label{sec:summary}

In this work we have proposed to generate quantum transport using cold
atoms in an optical superlattice that is modulated periodically both
in time and in space. Since there is no average force acting on the
atoms, we deal with a quantum ratchet effect that makes atoms on even
and odd sites move along opposite directions.

We have demonstrated that the ratchet effect can be induced both for
free particles and in the case of nonzero on-site interactions. The
latter represents to our knowledge the first time strong many-body
interactions have been treated exactly in Hamiltonian ratchets
\footnote{In the work of \cite{poletti07} the ratchet is studied using
  the mean-field formalism for a Bose-Einstein condensate.}

This ratchet effect can be used to transport quantum information by
imprinting a qubit or an entangled state on one or more atoms of the
atomic chain and letting the system evolve according to our ratchet
potentials. The dynamics generated with our scheme corresponds to a
smooth concatenation of perfect swap gates. Therefore, after a time $n
\times T$, we will find that the qubit state has been transported a
well determined distance, of order $n$, along the lattice.

Our ideas could be implemented in current experiments with optical
superlattices. From the quantum information point of view, compared to
other proposals based on effective spin interactions between atoms,
the ratchet effect should be faster and lead to a more flexible
dynamics. From the fundamental point of view, we believe that the
modulated superlattices are a rich playground in which to study
transport phenomena, quantum diffusion and the influence of noise and
of chaos in the transport of quantum states.

We thank J. Calsamiglia for useful discussions. We are grateful to
A. Sanpera, R. Munoz-Tapia, and A. Kay for a careful reading of the
manuscript. O.R.I. acknowledges support from the spanish MEC grants
AP2005-0595, FIS2005-03169, Consolider-Ingenio2010 CSD2006-00019 QOIT,
and Catalan grant SGR-00185. J.J.G.R acknowledges financial support
from the Ramon y Cajal Program and from the spanish projects
FIS2006-04885 and CAM-UCM/910758.


\begin{thebibliography}{51}
\expandafter\ifx\csname natexlab\endcsname\relax\def\natexlab#1{#1}\fi
\expandafter\ifx\csname bibnamefont\endcsname\relax
  \def\bibnamefont#1{#1}\fi
\expandafter\ifx\csname bibfnamefont\endcsname\relax
  \def\bibfnamefont#1{#1}\fi
\expandafter\ifx\csname citenamefont\endcsname\relax
  \def\citenamefont#1{#1}\fi
\expandafter\ifx\csname url\endcsname\relax
  \def\url#1{\texttt{#1}}\fi
\expandafter\ifx\csname urlprefix\endcsname\relax\def\urlprefix{URL }\fi
\providecommand{\bibinfo}[2]{#2}
\providecommand{\eprint}[2][]{\url{#2}}

\bibitem[{\citenamefont{{Reimann}}(1997)}]{reimann97}
\bibinfo{author}{\bibfnamefont{P.}~\bibnamefont{{Reimann}}},
  \bibinfo{journal}{Phys.~Rep.} \textbf{\bibinfo{volume}{290}},
  \bibinfo{pages}{149} (\bibinfo{year}{1997}).

\bibitem[{\citenamefont{Astumian}(1997)}]{astumian97}
\bibinfo{author}{\bibfnamefont{R.~D.} \bibnamefont{Astumian}},
  \bibinfo{journal}{Science} \textbf{\bibinfo{volume}{276}},
  \bibinfo{pages}{917} (\bibinfo{year}{1997}).

\bibitem[{\citenamefont{Flach et~al.}(2000)\citenamefont{Flach, Yevtushenko,
  and Zolotaryuk}}]{flach00}
\bibinfo{author}{\bibfnamefont{S.}~\bibnamefont{Flach}},
  \bibinfo{author}{\bibfnamefont{O.}~\bibnamefont{Yevtushenko}},
  \bibnamefont{and}
  \bibinfo{author}{\bibfnamefont{Y.}~\bibnamefont{Zolotaryuk}},
  \bibinfo{journal}{Phys. Rev. Lett.} \textbf{\bibinfo{volume}{84}},
  \bibinfo{pages}{2358} (\bibinfo{year}{2000}).

\bibitem[{\citenamefont{Schanz et~al.}(2001)\citenamefont{Schanz, Otto,
  Ketzmerick, and Dittrich}}]{schanz01}
\bibinfo{author}{\bibfnamefont{H.}~\bibnamefont{Schanz}},
  \bibinfo{author}{\bibfnamefont{M.~F.} \bibnamefont{Otto}},
  \bibinfo{author}{\bibfnamefont{R.}~\bibnamefont{Ketzmerick}},
  \bibnamefont{and} \bibinfo{author}{\bibfnamefont{T.}~\bibnamefont{Dittrich}},
  \bibinfo{journal}{Phys. Rev. Lett.} \textbf{\bibinfo{volume}{87}},
  \bibinfo{pages}{070601} (\bibinfo{year}{2001}).

\bibitem[{\citenamefont{Jones et~al.}(2004)\citenamefont{Jones, Goonasekera,
  and Renzoni}}]{jones04}
\bibinfo{author}{\bibfnamefont{P.~H.} \bibnamefont{Jones}},
  \bibinfo{author}{\bibfnamefont{M.}~\bibnamefont{Goonasekera}},
  \bibnamefont{and} \bibinfo{author}{\bibfnamefont{F.}~\bibnamefont{Renzoni}},
  \bibinfo{journal}{Phys. Rev. Lett.} \textbf{\bibinfo{volume}{93}},
  \bibinfo{pages}{073904} (\bibinfo{year}{2004}).

\bibitem[{\citenamefont{Moore et~al.}(1995)\citenamefont{Moore, Robinson,
  Bharucha, Sundaram, and Raizen}}]{moore95}
\bibinfo{author}{\bibfnamefont{F.~L.} \bibnamefont{Moore}},
  \bibinfo{author}{\bibfnamefont{J.~C.} \bibnamefont{Robinson}},
  \bibinfo{author}{\bibfnamefont{C.~F.} \bibnamefont{Bharucha}},
  \bibinfo{author}{\bibfnamefont{B.}~\bibnamefont{Sundaram}}, \bibnamefont{and}
  \bibinfo{author}{\bibfnamefont{M.~G.} \bibnamefont{Raizen}},
  \bibinfo{journal}{Phys. Rev. Lett.} \textbf{\bibinfo{volume}{75}},
  \bibinfo{pages}{4598} (\bibinfo{year}{1995}).

\bibitem[{\citenamefont{Klappauf et~al.}(1998)\citenamefont{Klappauf, Oskay,
  Steck, and Raizen}}]{klappauf98}
\bibinfo{author}{\bibfnamefont{B.~G.} \bibnamefont{Klappauf}},
  \bibinfo{author}{\bibfnamefont{W.~H.} \bibnamefont{Oskay}},
  \bibinfo{author}{\bibfnamefont{D.~A.} \bibnamefont{Steck}}, \bibnamefont{and}
  \bibinfo{author}{\bibfnamefont{M.~G.} \bibnamefont{Raizen}},
  \bibinfo{journal}{Phys. Rev. Lett.} \textbf{\bibinfo{volume}{81}},
  \bibinfo{pages}{1203} (\bibinfo{year}{1998}).

\bibitem[{\citenamefont{Mennerat-Robilliard
  et~al.}(1999)\citenamefont{Mennerat-Robilliard, Lucas, Guibal, Tabosa,
  Jurczak, Courtois, and Grynberg}}]{mennerat99}
\bibinfo{author}{\bibfnamefont{C.}~\bibnamefont{Mennerat-Robilliard}},
  \bibinfo{author}{\bibfnamefont{D.}~\bibnamefont{Lucas}},
  \bibinfo{author}{\bibfnamefont{S.}~\bibnamefont{Guibal}},
  \bibinfo{author}{\bibfnamefont{J.}~\bibnamefont{Tabosa}},
  \bibinfo{author}{\bibfnamefont{C.}~\bibnamefont{Jurczak}},
  \bibinfo{author}{\bibfnamefont{J.-Y.} \bibnamefont{Courtois}},
  \bibnamefont{and} \bibinfo{author}{\bibfnamefont{G.}~\bibnamefont{Grynberg}},
  \bibinfo{journal}{Phys. Rev. Lett.} \textbf{\bibinfo{volume}{82}},
  \bibinfo{pages}{851} (\bibinfo{year}{1999}).

\bibitem[{\citenamefont{Ringot et~al.}(2000)\citenamefont{Ringot, Szriftgiser,
  Garreau, and Delande}}]{ringot00}
\bibinfo{author}{\bibfnamefont{J.}~\bibnamefont{Ringot}},
  \bibinfo{author}{\bibfnamefont{P.}~\bibnamefont{Szriftgiser}},
  \bibinfo{author}{\bibfnamefont{J.~C.} \bibnamefont{Garreau}},
  \bibnamefont{and} \bibinfo{author}{\bibfnamefont{D.}~\bibnamefont{Delande}},
  \bibinfo{journal}{Phys. Rev. Lett.} \textbf{\bibinfo{volume}{85}},
  \bibinfo{pages}{2741} (\bibinfo{year}{2000}).

\bibitem[{\citenamefont{Behinaein et~al.}(2006)\citenamefont{Behinaein,
  Ramareddy, Ahmadi, and Summy}}]{behinaein06}
\bibinfo{author}{\bibfnamefont{G.}~\bibnamefont{Behinaein}},
  \bibinfo{author}{\bibfnamefont{V.}~\bibnamefont{Ramareddy}},
  \bibinfo{author}{\bibfnamefont{P.}~\bibnamefont{Ahmadi}}, \bibnamefont{and}
  \bibinfo{author}{\bibfnamefont{G.~S.} \bibnamefont{Summy}},
  \bibinfo{journal}{Phys. Rev. Lett.} \textbf{\bibinfo{volume}{97}},
  \bibinfo{pages}{244101} (\bibinfo{year}{2006}).

\bibitem[{\citenamefont{Gommers et~al.}(2006)\citenamefont{Gommers, Denisov,
  and Renzoni}}]{gommers06}
\bibinfo{author}{\bibfnamefont{R.}~\bibnamefont{Gommers}},
  \bibinfo{author}{\bibfnamefont{S.}~\bibnamefont{Denisov}}, \bibnamefont{and}
  \bibinfo{author}{\bibfnamefont{F.}~\bibnamefont{Renzoni}},
  \bibinfo{journal}{Phys. Rev. Lett.} \textbf{\bibinfo{volume}{96}},
  \bibinfo{pages}{240604} (\bibinfo{year}{2006}).

\bibitem[{\citenamefont{Sanchez-Palencia}(2004)}]{sanchez-palencia04}
\bibinfo{author}{\bibfnamefont{L.}~\bibnamefont{Sanchez-Palencia}},
  \bibinfo{journal}{Phys. Rev. E} \textbf{\bibinfo{volume}{70}}, \bibinfo{eid}{011102}
  (\bibinfo{year}{2004}).

\bibitem[{\citenamefont{Sjolund et~al.}(2006)\citenamefont{Sjolund, Petra,
  Dion, Jonsell, Nylen, Sanchez-Palencia, and Kastberg}}]{sjolund06}
\bibinfo{author}{\bibfnamefont{P.}~\bibnamefont{Sjolund}},
  \bibinfo{author}{\bibfnamefont{S.~J.~H.} \bibnamefont{Petra}},
  \bibinfo{author}{\bibfnamefont{C.~M.} \bibnamefont{Dion}},
  \bibinfo{author}{\bibfnamefont{S.}~\bibnamefont{Jonsell}},
  \bibinfo{author}{\bibfnamefont{M.}~\bibnamefont{Nylen}},
  \bibinfo{author}{\bibfnamefont{L.}~\bibnamefont{Sanchez-Palencia}},
  \bibnamefont{and} \bibinfo{author}{\bibfnamefont{A.}~\bibnamefont{Kastberg}},
  \bibinfo{journal}{Phys. Rev. Let..} \textbf{\bibinfo{volume}{96}},
  \bibinfo{eid}{190602} (\bibinfo{year}{2006}).
  
  
  \bibitem[{\citenamefont{Denisov et~al.}(2007)\citenamefont{S. Denisov and L. Morales-Molina and S. Flach and P. H\"{a}nggi}}]{denisov07}
\bibinfo{author}{\bibfnamefont{S.}~\bibnamefont{Denisov}},
  \bibinfo{author}{\bibfnamefont{L.} \bibnamefont{Morales-Molina}},
  \bibinfo{author}{\bibfnamefont{S.} \bibnamefont{Flach}},
    \bibinfo{author}{\bibfnamefont{P.} \bibnamefont{ H\"{a}nggi}},
  \bibinfo{journal}{Phys. Rev. A} \textbf{\bibinfo{volume}{75}}, \bibinfo{eid}{063424}
  (\bibinfo{year}{2007}).

\bibitem[{\citenamefont{Poletti et~al.}(2007)\citenamefont{Poletti, Benenti,
  Casati, and Li}}]{poletti07}
\bibinfo{author}{\bibfnamefont{D.}~\bibnamefont{Poletti}},
  \bibinfo{author}{\bibfnamefont{G.}~\bibnamefont{Benenti}},
  \bibinfo{author}{\bibfnamefont{G.}~\bibnamefont{Casati}}, \bibnamefont{and}
  \bibinfo{author}{\bibfnamefont{B.}~\bibnamefont{Li}},
  \bibinfo{journal}{Phys. Rev. A}
  \textbf{\bibinfo{volume}{76}}, \bibinfo{eid}{023421}
   (\bibinfo{year}{2007}).

\bibitem[{\citenamefont{Monteiro et~al.}(2002)\citenamefont{Monteiro, Dando,
  Hutchings, and Isherwood}}]{monteiro02}
\bibinfo{author}{\bibfnamefont{T.~S.} \bibnamefont{Monteiro}},
  \bibinfo{author}{\bibfnamefont{P.~A.} \bibnamefont{Dando}},
  \bibinfo{author}{\bibfnamefont{N.~A.~C.} \bibnamefont{Hutchings}},
  \bibnamefont{and} \bibinfo{author}{\bibfnamefont{M.~R.}
  \bibnamefont{Isherwood}}, \bibinfo{journal}{Phys. Rev. Lett.}
  \textbf{\bibinfo{volume}{89}}, \bibinfo{pages}{194102}
  (\bibinfo{year}{2002}).

\bibitem[{\citenamefont{Jonckheere et~al.}(2003)\citenamefont{Jonckheere,
  Isherwood, and Monteiro}}]{jonckheere03}
\bibinfo{author}{\bibfnamefont{T.}~\bibnamefont{Jonckheere}},
  \bibinfo{author}{\bibfnamefont{M.~R.} \bibnamefont{Isherwood}},
  \bibnamefont{and} \bibinfo{author}{\bibfnamefont{T.~S.}
  \bibnamefont{Monteiro}}, \bibinfo{journal}{Phys. Rev. Lett.}
  \textbf{\bibinfo{volume}{91}}, \bibinfo{pages}{253003}
  (\bibinfo{year}{2003}).

\bibitem[{\citenamefont{Hutchings et~al.}(2004)\citenamefont{Hutchings,
  Isherwood, Jonckheere, and Monteiro}}]{hutchings04}
\bibinfo{author}{\bibfnamefont{N.~A.~C.} \bibnamefont{Hutchings}},
  \bibinfo{author}{\bibfnamefont{M.~R.} \bibnamefont{Isherwood}},
  \bibinfo{author}{\bibfnamefont{T.}~\bibnamefont{Jonckheere}},
  \bibnamefont{and} \bibinfo{author}{\bibfnamefont{T.~S.}
  \bibnamefont{Monteiro}}, \bibinfo{journal}{Phys. Rev. E}
  \textbf{\bibinfo{volume}{70}}, \bibinfo{pages}{036205}
  (\bibinfo{year}{2004}).

\bibitem[{\citenamefont{Hur et~al.}(2005)\citenamefont{Hur, Creffield, Jones,
  and Monteiro}}]{hur05}
\bibinfo{author}{\bibfnamefont{G.}~\bibnamefont{Hur}},
  \bibinfo{author}{\bibfnamefont{C.~E.} \bibnamefont{Creffield}},
  \bibinfo{author}{\bibfnamefont{P.~H.} \bibnamefont{Jones}}, \bibnamefont{and}
  \bibinfo{author}{\bibfnamefont{T.~S.} \bibnamefont{Monteiro}},
  \bibinfo{journal}{Phys. Rev. A} \textbf{\bibinfo{volume}{72}},
  \bibinfo{pages}{013403} (\bibinfo{year}{2005}).

\bibitem[{\citenamefont{Kenfack et~al.}(2007)\citenamefont{Kenfack, Gong, and
  Pattanayak}}]{kenfack07}
\bibinfo{author}{\bibfnamefont{A.}~\bibnamefont{Kenfack}},
  \bibinfo{author}{\bibfnamefont{J.}~\bibnamefont{Gong}}, \bibnamefont{and}
  \bibinfo{author}{\bibfnamefont{A.~K.} \bibnamefont{Pattanayak}},
  \bibinfo{journal}{arXiv:0708.3026}  (\bibinfo{year}{2007}).

\bibitem[{\citenamefont{Bennett and DiVincenzo}(2000)}]{bennett00}
\bibinfo{author}{\bibfnamefont{C.~H.} \bibnamefont{Bennett}} \bibnamefont{and}
  \bibinfo{author}{\bibfnamefont{D.~P.} \bibnamefont{DiVincenzo}},
  \bibinfo{journal}{Nature} \textbf{\bibinfo{volume}{404}},
  \bibinfo{pages}{247} (\bibinfo{year}{2000}).

\bibitem[{\citenamefont{Bennett et~al.}(1993)\citenamefont{Bennett, Brassard,
  Crepeau, Jozsa, Peres, and Wootters}}]{bennett93}
\bibinfo{author}{\bibfnamefont{C.~H.} \bibnamefont{Bennett}},
  \bibinfo{author}{\bibfnamefont{G.}~\bibnamefont{Brassard}},
  \bibinfo{author}{\bibfnamefont{C.}~\bibnamefont{Crepeau}},
  \bibinfo{author}{\bibfnamefont{R.}~\bibnamefont{Jozsa}},
  \bibinfo{author}{\bibfnamefont{A.}~\bibnamefont{Peres}}, \bibnamefont{and}
  \bibinfo{author}{\bibfnamefont{W.~K.} \bibnamefont{Wootters}},
  \bibinfo{journal}{Phys. Rev. Lett.} \textbf{\bibinfo{volume}{70}},
  \bibinfo{pages}{1895} (\bibinfo{year}{1993}).

\bibitem[{\citenamefont{Briegel et~al.}(1998)\citenamefont{Briegel, D\"ur,
  Cirac, and Zoller}}]{briegel98}
\bibinfo{author}{\bibfnamefont{H.-J.} \bibnamefont{Briegel}},
  \bibinfo{author}{\bibfnamefont{W.}~\bibnamefont{D\"ur}},
  \bibinfo{author}{\bibfnamefont{J.~I.} \bibnamefont{Cirac}}, \bibnamefont{and}
  \bibinfo{author}{\bibfnamefont{P.}~\bibnamefont{Zoller}},
  \bibinfo{journal}{Phys. Rev. Lett.} \textbf{\bibinfo{volume}{81}},
  \bibinfo{pages}{5932} (\bibinfo{year}{1998}).

\bibitem[{\citenamefont{Duan et~al.}(2001)\citenamefont{Duan, Lukin, Cirac, and
  Zoller}}]{duan01}
\bibinfo{author}{\bibfnamefont{L.-M.} \bibnamefont{Duan}},
  \bibinfo{author}{\bibfnamefont{M.~D.} \bibnamefont{Lukin}},
  \bibinfo{author}{\bibfnamefont{J.~I.} \bibnamefont{Cirac}}, \bibnamefont{and}
  \bibinfo{author}{\bibfnamefont{P.}~\bibnamefont{Zoller}},
  \bibinfo{journal}{Nature} \textbf{\bibinfo{volume}{414}},
  \bibinfo{pages}{413} (\bibinfo{year}{2001}).

\bibitem[{\citenamefont{Bose}(2003)}]{bose03}
\bibinfo{author}{\bibfnamefont{S.}~\bibnamefont{Bose}}, \bibinfo{journal}{Phys.
  Rev. Lett.} \textbf{\bibinfo{volume}{91}}, \bibinfo{pages}{207901}
  (\bibinfo{year}{2003}).

\bibitem[{\citenamefont{Osborne and Linden}(2004)}]{osborne04}
\bibinfo{author}{\bibfnamefont{T.~J.} \bibnamefont{Osborne}} \bibnamefont{and}
  \bibinfo{author}{\bibfnamefont{N.}~\bibnamefont{Linden}},
  \bibinfo{journal}{Phys. Rev. A} \textbf{\bibinfo{volume}{69}},
  \bibinfo{pages}{052315} (\bibinfo{year}{2004}).

\bibitem[{\citenamefont{Amico et~al.}(2004)\citenamefont{Amico, Osterloh,
  Plastina, Fazio, and Palma}}]{amico04}
\bibinfo{author}{\bibfnamefont{L.}~\bibnamefont{Amico}},
  \bibinfo{author}{\bibfnamefont{A.}~\bibnamefont{Osterloh}},
  \bibinfo{author}{\bibfnamefont{F.}~\bibnamefont{Plastina}},
  \bibinfo{author}{\bibfnamefont{R.}~\bibnamefont{Fazio}}, \bibnamefont{and}
  \bibinfo{author}{\bibfnamefont{G.~M.} \bibnamefont{Palma}},
  \bibinfo{journal}{Phys. Rev. A} \textbf{\bibinfo{volume}{69}},
  \bibinfo{pages}{022304} (\bibinfo{year}{2004}).

\bibitem[{\citenamefont{Christandl et~al.}(2004)\citenamefont{Christandl,
  Datta, Ekert, and Landahl}}]{christandl04}
\bibinfo{author}{\bibfnamefont{M.}~\bibnamefont{Christandl}},
  \bibinfo{author}{\bibfnamefont{N.}~\bibnamefont{Datta}},
  \bibinfo{author}{\bibfnamefont{A.}~\bibnamefont{Ekert}}, \bibnamefont{and}
  \bibinfo{author}{\bibfnamefont{A.~J.} \bibnamefont{Landahl}},
  \bibinfo{journal}{Phys. Rev. Lett.} \textbf{\bibinfo{volume}{92}},
  \bibinfo{pages}{187902} (\bibinfo{year}{2004}).

\bibitem[{\citenamefont{Plenio and Semiao}(2005)}]{plenio04}
\bibinfo{author}{\bibfnamefont{M.~B.} \bibnamefont{Plenio}} \bibnamefont{and}
  \bibinfo{author}{\bibfnamefont{F.~L.} \bibnamefont{Semiao}},
  \bibinfo{journal}{New J. Phys.} \textbf{\bibinfo{volume}{7}},
  \bibinfo{pages}{73} (\bibinfo{year}{2005}).

\bibitem[{\citenamefont{Wojcik et~al.}(2005)\citenamefont{Wojcik, Luczak,
  Kurzynski, Grudka, Gdala, and Bednarska}}]{wojcik05}
\bibinfo{author}{\bibfnamefont{A.}~\bibnamefont{Wojcik}},
  \bibinfo{author}{\bibfnamefont{T.}~\bibnamefont{Luczak}},
  \bibinfo{author}{\bibfnamefont{P.}~\bibnamefont{Kurzynski}},
  \bibinfo{author}{\bibfnamefont{A.}~\bibnamefont{Grudka}},
  \bibinfo{author}{\bibfnamefont{T.}~\bibnamefont{Gdala}}, \bibnamefont{and}
  \bibinfo{author}{\bibfnamefont{M.}~\bibnamefont{Bednarska}},
  \bibinfo{journal}{Phys. Rev. A} \textbf{\bibinfo{volume}{72}},
  \bibinfo{pages}{034303} (\bibinfo{year}{2005}).

\bibitem[{\citenamefont{Haselgrove}(2005)}]{haselgrove05}
\bibinfo{author}{\bibfnamefont{H.~L.} \bibnamefont{Haselgrove}},
  \bibinfo{journal}{Phys. Rev. A} \textbf{\bibinfo{volume}{72}},
  \bibinfo{pages}{062326} (\bibinfo{year}{2005}).

\bibitem[{\citenamefont{Burgarth and Bose}(2005)}]{burgarth05}
\bibinfo{author}{\bibfnamefont{D.}~\bibnamefont{Burgarth}} \bibnamefont{and}
  \bibinfo{author}{\bibfnamefont{S.}~\bibnamefont{Bose}},
  \bibinfo{journal}{Phys. Rev. A} \textbf{\bibinfo{volume}{71}},
  \bibinfo{pages}{052315} (\bibinfo{year}{2005}).

\bibitem[{\citenamefont{Burgarth}(2007)}]{burgarth07}
\bibinfo{author}{\bibfnamefont{D.}~\bibnamefont{Burgarth}},
  \bibinfo{journal}{PhD Thesis, arXiv:0704.1309}  (\bibinfo{year}{2007}).

\bibitem[{\citenamefont{Romero-Isart
  et~al.}(2007{\natexlab{a}})\citenamefont{Romero-Isart, Eckert, and
  Sanpera}}]{romero-isart07}
\bibinfo{author}{\bibfnamefont{O.}~\bibnamefont{Romero-Isart}},
  \bibinfo{author}{\bibfnamefont{K.}~\bibnamefont{Eckert}}, \bibnamefont{and}
  \bibinfo{author}{\bibfnamefont{A.}~\bibnamefont{Sanpera}},
  \bibinfo{journal}{Phys. Rev. A} \textbf{\bibinfo{volume}{75}},
  \bibinfo{pages}{050303} (\bibinfo{year}{2007}{\natexlab{a}}).

\bibitem[{\citenamefont{Eckert et~al.}(2007)\citenamefont{Eckert, Romero-Isart,
  and Sanpera}}]{eckert07}
\bibinfo{author}{\bibfnamefont{K.}~\bibnamefont{Eckert}},
  \bibinfo{author}{\bibfnamefont{O.}~\bibnamefont{Romero-Isart}},
  \bibnamefont{and} \bibinfo{author}{\bibfnamefont{A.}~\bibnamefont{Sanpera}},
  \bibinfo{journal}{New J. Physics} \textbf{\bibinfo{volume}{9}},
  \bibinfo{pages}{155} (\bibinfo{year}{2007}).

\bibitem[{\citenamefont{Bayat and Bose}(2007)}]{bayat07}
\bibinfo{author}{\bibfnamefont{A.}~\bibnamefont{Bayat}} \bibnamefont{and}
  \bibinfo{author}{\bibfnamefont{S.}~\bibnamefont{Bose}},
  \bibinfo{journal}{arXiv:0706.4176}  (\bibinfo{year}{2007}).

\bibitem[{\citenamefont{Kay}(2007)}]{kay07}
\bibinfo{author}{\bibfnamefont{A.}~\bibnamefont{Kay}}, \bibinfo{journal}{Phys.
  Rev. Lett.} \textbf{\bibinfo{volume}{98}}, \bibinfo{pages}{010501}
  (\bibinfo{year}{2007}).

\bibitem[{\citenamefont{D'Amico et~al.}(2007)\citenamefont{D'Amico, Lovett, and
  Spiller}}]{damico07}
\bibinfo{author}{\bibfnamefont{I.}~\bibnamefont{D'Amico}},
  \bibinfo{author}{\bibfnamefont{B.~W.} \bibnamefont{Lovett}},
  \bibnamefont{and} \bibinfo{author}{\bibfnamefont{T.~P.}
  \bibnamefont{Spiller}}, \bibinfo{journal}{Phys. Rev. A}
  \textbf{\bibinfo{volume}{76}}, \bibinfo{pages}{030302(R)}
  (\bibinfo{year}{2007}).

\bibitem[{\citenamefont{Romero-Isart
  et~al.}(2007{\natexlab{b}})\citenamefont{Romero-Isart, Eckert, Rod{\'o}, and
  Sanpera}}]{romeroisart07b}
\bibinfo{author}{\bibfnamefont{O.}~\bibnamefont{Romero-Isart}},
  \bibinfo{author}{\bibfnamefont{K.}~\bibnamefont{Eckert}},
  \bibinfo{author}{\bibfnamefont{C.}~\bibnamefont{Rod{\'o}}}, \bibnamefont{and}
  \bibinfo{author}{\bibfnamefont{A.}~\bibnamefont{Sanpera}},
  \bibinfo{journal}{J. Phys. A: Math. Theor.} \textbf{\bibinfo{volume}{40}},
  \bibinfo{pages}{8019} (\bibinfo{year}{2007}{\natexlab{b}}).

\bibitem[{\citenamefont{Bose}(2006)}]{bose07}
\bibinfo{author}{\bibfnamefont{S.}~\bibnamefont{Bose}},
  \bibinfo{journal}{cond-mat/0610024}  (\bibinfo{year}{2006}).

\bibitem[{\citenamefont{F\"olling et~al.}(2007)\citenamefont{F\"olling,
  Trotzky, Cheinet, Feld, Saers, Widera, M\"uller, and Bloch}}]{foelling07}
\bibinfo{author}{\bibfnamefont{S.}~\bibnamefont{F\"olling}},
  \bibinfo{author}{\bibfnamefont{S.}~\bibnamefont{Trotzky}},
  \bibinfo{author}{\bibfnamefont{P.}~\bibnamefont{Cheinet}},
  \bibinfo{author}{\bibfnamefont{M.}~\bibnamefont{Feld}},
  \bibinfo{author}{\bibfnamefont{R.}~\bibnamefont{Saers}},
  \bibinfo{author}{\bibfnamefont{A.}~\bibnamefont{Widera}},
  \bibinfo{author}{\bibfnamefont{T.}~\bibnamefont{M\"uller}}, \bibnamefont{and}
  \bibinfo{author}{\bibfnamefont{I.}~\bibnamefont{Bloch}},
  \bibinfo{journal}{Nature} \textbf{\bibinfo{volume}{448}},
  \bibinfo{pages}{1029} (\bibinfo{year}{2007}).

\bibitem[{\citenamefont{Anderlini et~al.}(2007)\citenamefont{Anderlini, Lee,
  Brown, Sebby-Strabley, Phillips, and Porto}}]{anderlini07}
\bibinfo{author}{\bibfnamefont{M.}~\bibnamefont{Anderlini}},
  \bibinfo{author}{\bibfnamefont{P.~J.} \bibnamefont{Lee}},
  \bibinfo{author}{\bibfnamefont{B.~L.} \bibnamefont{Brown}},
  \bibinfo{author}{\bibfnamefont{J.}~\bibnamefont{Sebby-Strabley}},
  \bibinfo{author}{\bibfnamefont{W.~D.} \bibnamefont{Phillips}},
  \bibnamefont{and} \bibinfo{author}{\bibfnamefont{J.~V.} \bibnamefont{Porto}},
  \bibinfo{journal}{Nature} \textbf{\bibinfo{volume}{448}},
  \bibinfo{pages}{452} (\bibinfo{year}{2007}).

\bibitem[{\citenamefont{Lee et~al.}(2007)\citenamefont{Lee, Anderlini, Brown,
  Sebby-Strabley, Phillips, and Porto}}]{lee07}
\bibinfo{author}{\bibfnamefont{P.~J.} \bibnamefont{Lee}},
  \bibinfo{author}{\bibfnamefont{M.}~\bibnamefont{Anderlini}},
  \bibinfo{author}{\bibfnamefont{B.~L.} \bibnamefont{Brown}},
  \bibinfo{author}{\bibfnamefont{J.}~\bibnamefont{Sebby-Strabley}},
  \bibinfo{author}{\bibfnamefont{W.~D.} \bibnamefont{Phillips}},
  \bibnamefont{and} \bibinfo{author}{\bibfnamefont{J.~V.} \bibnamefont{Porto}},
  \bibinfo{journal}{Phys. Rev. Lett.} \textbf{\bibinfo{volume}{99}},
  \bibinfo{pages}{020402} (\bibinfo{year}{2007}).

\bibitem[{\citenamefont{{Anderlini} et~al.}(2006)\citenamefont{{Anderlini},
  {Sebby-Strabley}, {Kruse}, {Porto}, and {Phillips}}}]{anderlini06}
\bibinfo{author}{\bibfnamefont{M.}~\bibnamefont{{Anderlini}}},
  \bibinfo{author}{\bibfnamefont{J.}~\bibnamefont{{Sebby-Strabley}}},
  \bibinfo{author}{\bibfnamefont{J.}~\bibnamefont{{Kruse}}},
  \bibinfo{author}{\bibfnamefont{J.~V.} \bibnamefont{{Porto}}},
  \bibnamefont{and} \bibinfo{author}{\bibfnamefont{W.~D.}
  \bibnamefont{{Phillips}}}, \bibinfo{journal}{J. Phys. B: At. Mol. Op.}
  \textbf{\bibinfo{volume}{39}}, \bibinfo{pages}{199} (\bibinfo{year}{2006}).

\bibitem[{\citenamefont{{Sebby-Strabley}
  et~al.}(2006)\citenamefont{{Sebby-Strabley}, {Anderlini}, {Jessen}, and
  {Porto}}}]{sebbystrabley06}
\bibinfo{author}{\bibfnamefont{J.}~\bibnamefont{{Sebby-Strabley}}},
  \bibinfo{author}{\bibfnamefont{M.}~\bibnamefont{{Anderlini}}},
  \bibinfo{author}{\bibfnamefont{P.~S.} \bibnamefont{{Jessen}}},
  \bibnamefont{and} \bibinfo{author}{\bibfnamefont{J.~V.}
  \bibnamefont{{Porto}}}, \bibinfo{journal}{Phys. Rev. A}
  \textbf{\bibinfo{volume}{73}}, \bibinfo{pages}{033605}
  (\bibinfo{year}{2006}).

\bibitem[{\citenamefont{{Jaksch} et~al.}(1998)\citenamefont{{Jaksch}, {Bruder},
  {Cirac}, {Gardiner}, and {Zoller}}}]{jaksch98}
\bibinfo{author}{\bibfnamefont{D.}~\bibnamefont{{Jaksch}}},
  \bibinfo{author}{\bibfnamefont{C.}~\bibnamefont{{Bruder}}},
  \bibinfo{author}{\bibfnamefont{J.~I.} \bibnamefont{{Cirac}}},
  \bibinfo{author}{\bibfnamefont{C.~W.} \bibnamefont{{Gardiner}}},
  \bibnamefont{and} \bibinfo{author}{\bibfnamefont{P.}~\bibnamefont{{Zoller}}},
  \bibinfo{journal}{Phys. Rev. Lett.} \textbf{\bibinfo{volume}{81}},
  \bibinfo{pages}{3108} (\bibinfo{year}{1998}).

\bibitem[{\citenamefont{{Greiner} et~al.}(2002)\citenamefont{{Greiner},
  {Mandel}, {Esslinger}, {H{\"a}nsch}, and {Bloch}}}]{greiner02}
\bibinfo{author}{\bibfnamefont{M.}~\bibnamefont{{Greiner}}},
  \bibinfo{author}{\bibfnamefont{O.}~\bibnamefont{{Mandel}}},
  \bibinfo{author}{\bibfnamefont{T.}~\bibnamefont{{Esslinger}}},
  \bibinfo{author}{\bibfnamefont{T.~W.} \bibnamefont{{H{\"a}nsch}}},
  \bibnamefont{and} \bibinfo{author}{\bibfnamefont{I.}~\bibnamefont{{Bloch}}},
  \bibinfo{journal}{Nature} \textbf{\bibinfo{volume}{415}}, \bibinfo{pages}{39}
  (\bibinfo{year}{2002}).

\bibitem[{\citenamefont{Khaneja et~al.}(2001)\citenamefont{Khaneja, Brockett,
  and Glaser}}]{khaneja01}
\bibinfo{author}{\bibfnamefont{N.}~\bibnamefont{Khaneja}},
  \bibinfo{author}{\bibfnamefont{R.}~\bibnamefont{Brockett}}, \bibnamefont{and}
  \bibinfo{author}{\bibfnamefont{S.~J.} \bibnamefont{Glaser}},
  \bibinfo{journal}{Phys. Rev. A} \textbf{\bibinfo{volume}{63}},
  \bibinfo{pages}{032308} (\bibinfo{year}{2001}).

\bibitem[{\citenamefont{Branch et~al.}(1999)\citenamefont{Branch, Coleman, and
  Li}}]{branch99}
\bibinfo{author}{\bibfnamefont{M.~A.} \bibnamefont{Branch}},
  \bibinfo{author}{\bibfnamefont{T.~F.} \bibnamefont{Coleman}},
  \bibnamefont{and} \bibinfo{author}{\bibfnamefont{Y.}~\bibnamefont{Li}},
  \bibinfo{journal}{SIAM J. Sci. Comput.} \textbf{\bibinfo{volume}{21}},
  \bibinfo{pages}{1} (\bibinfo{year}{1999}).

\bibitem[{\citenamefont{F\"olling et~al.}(2006)\citenamefont{F\"olling, Widera,
  Muller, Gerbier, and Bloch}}]{foelling06}
\bibinfo{author}{\bibfnamefont{S.}~\bibnamefont{F\"olling}},
  \bibinfo{author}{\bibfnamefont{A.}~\bibnamefont{Widera}},
  \bibinfo{author}{\bibfnamefont{T.}~\bibnamefont{Muller}},
  \bibinfo{author}{\bibfnamefont{F.}~\bibnamefont{Gerbier}}, \bibnamefont{and}
  \bibinfo{author}{\bibfnamefont{I.}~\bibnamefont{Bloch}},
  \bibinfo{journal}{Phys. Rev. Lett.} \textbf{\bibinfo{volume}{97}},
  \bibinfo{pages}{060403} (\bibinfo{year}{2006}).

\bibitem[{\citenamefont{Werschnik and Gross}()}]{werschnik07}
\bibinfo{author}{\bibfnamefont{J.}~\bibnamefont{Werschnik}} \bibnamefont{and}
  \bibinfo{author}{\bibfnamefont{E.~K.~U.} \bibnamefont{Gross}},
  \bibinfo{note}{arXiv:0707.1883v1}.

\bibitem[{\citenamefont{Creffield}(2007)}]{creffield07}
\bibinfo{author}{\bibfnamefont{C.~E.} \bibnamefont{Creffield}},
  \bibinfo{journal}{Phys. Rev. Lett.} \textbf{\bibinfo{volume}{99}},
  \bibinfo{pages}{110501} (\bibinfo{year}{2007}).
\end{thebibliography}
\end{document}